\centerline{\bf Can Naked Singularities Yield Gamma Ray Bursts?}
\bigskip
\centerline{H. M.  Antia}
\centerline{Tata Institute of Fundamental Research, Homi Bhabha Road, Mumbai
400 005, India
}
\centerline{email: antia@tifr.res.in}
\vskip 2 cm
{\narrower

\centerline{\bf Abstract}
\noindent
Gamma-ray bursts are believed to be the most luminous objects in the Universe.
There has been some suggestion that these arise from quantum
processes around naked singularities.
The main problem with this suggestion is that all known examples of
naked singularities are massless and hence there is effectively no
source of energy.
It is argued that a globally naked singularity coupled with quantum processes
operating within a distance of the order of Planck length of the singularity
will probably yield energy burst of the order of
$M_pc^2\approx2\times 10^{16}$ ergs, where $M_p$ is the Planck mass.
\bigskip
}

\noindent
Gamma-ray bursts are non-thermal bursts of $\gamma$-rays, believed to
have a total energy greater than $10^{51}$ ergs. The traditional models
involving merger of two neutron stars or collapse of a single
star into a black hole have some difficulty in explaining the amount
of energy emitted as the required efficiency in converting rest mass
energy (approximately $10^{54}$ ergs) to thermal energy may not be
achieved.
Witten~(1992) and Chakrabarti and Joshi~(1994) have suggested that these
bursts may arise from naked singularities.
Recently Singh~(1998) has claimed that efficiency of conversion of
gravitational energy to thermal energy can be increased if
the collapse is assumed to end up in naked singularity instead of
black hole. However, he has neither estimated the amount of energy
nor the time duration over which the energy is expected to be released in
such a process. 
In fact, their calculations (Barve et al. 1998) yield
a diverging flux and hence it is clear that their formulation is not
applicable for calculating the energy released by a naked singularity.
Divergence of flux in their calculation does not imply that such objects
would emit a large amount of energy.
It may be noted that in principle, there is no difficulty in generating
the required energy if the merging of two neutron stars leads to a black-hole.
The problem arises only when detailed calculations are done to estimate
the likely energy release. Thus, it is essential to do similar exercise for
naked singularities also and it would be interesting to
obtain a crude estimate of energy that can be released by such processes.

It may be recalled that naked singularities are gravitational
singularities which are not covered by a horizon.
Naked singularities can be either locally naked or globally naked.
From a locally naked singularity although the light rays can come out of
the singularity, they will fall back into the singularity and do not reach a
far away observer. While a globally naked singularity may be visible to an
observer at large distances. It is clear that for Gamma-ray bursts to be
observable at large distances we will need globally naked singularities.
It is by no means clear if globally naked singularities can form in generic
conditions with reasonable matter (Wald 1997; Brady et al.~1998), but
it is beyond our scope to discuss this issue.
Even if naked singularities can form in nature,
just because light rays can escape from a naked singularity does not imply
that it will produce a large burst of energy. For that one
also needs some source of energy.  All known examples of
naked singularities are massless (Singh 1996) and it is not clear how they
can generate significant energy. The energy can only come from surrounding
material and Singh~(1998) has not explained how he expects sufficient matter
to be present within the required distance from the singularity.
This factor will be crucial in determining the efficiency or viability of
the proposed model.

Singh~(1998) has suggested that quantum processes occurring near a
naked singularity
which is visible to far away observer can produce the energy required for
Gamma-ray bursts. The details of how the energy generation takes place have
not been worked  out and it is not clear how this energy can be computed.
Hence, we will attempt to estimate the amount
of energy that can possibly be emitted from a globally naked singularity
based on the amount of matter available as the source of energy, without
worrying about how this energy is actually converted to the required form.
Such estimates may be uncertain by a few orders of magnitude.
It is believed that quantum processes would generally operate
in a region which is within a distance of order of Planck length
($L_p=\sqrt{G\hbar/c^3}\approx10^{-33}$ cm), from the singularity.
Thus only the matter which is present in this volume can be expected
to be converted to
energy. In principle, by adjusting the initial conditions leading to
collapse it may be possible to bring arbitrarily large mass within such
a volume but such singularities will not be globally naked
even if they are locally naked. This means that even though the singularity
is locally naked, it is covered by a horizon at a finite distance from
the singularity and hence is not visible to observers outside the horizon.
This implies that energy generated by
such singularity will not reach far away observer.
The limiting mass in the vicinity of a globally naked singularity will
be of the order of $L_p c^2/G$, which is the Planck Mass
($M_p=\sqrt{\hbar c/G}\approx 2\times10^{-5}$ gm).
Thus the energy generated by a globally naked singularity would be of the
order of $M_pc^2\approx 2\times 10^{16}$ ergs.

In order to understand this let us consider spherically symmetric collapse
which is the only situation that has been worked out in any detail.
In this case (ignoring the shell crossing singularities)
 the singularity will form at the center and
let us assume that somehow 0.1 gm of matter ($5000 M_p$) has been
accumulated in the region within a distance of the order of $L_p$
from the singularity. Now this will form a
horizon at a distance of $2Gm/c^2\approx 10^{-29}$ cm. Thus whatever
energy is radiated from such a singularity will be confined inside this
horizon and will not be visible to a far away observer. In order to ensure
that the energy reaches a far away observer it will be necessary to ensure
that the singularity is also globally naked, which is not possible unless
the mass contained in this volume is much smaller. This gives an upper
limit on the amount of mass that can be accumulated in the vicinity of globally
naked singularity. Now 0.1 gm of matter will
at most generate $10^{20}$ ergs of energy, which appears to be an upper
limit to the energy that can be produced by a globally naked (shell
focusing) singularity in spherically symmetric collapse.

It may be noted that the only assumption
that is required in obtaining this limit is that only the mass which is
present within the distance of the order of $L_p$ from the singularity is
involved in energy generation. Further, the limiting energy is directly
proportional to the distance where the process is effective. Thus unless
this range is substantially increased by about 30 orders of magnitude
one cannot expect significant energy emission from a globally naked
singularity. In order to produce the required energy output, the quantum
process should operate over the distance of the order of Schwarzschild
radius, which is most unlikely.
Of course, there will be additional energy generated by the matter outside
through the normal process of compression etc., which will most probably be
larger than what comes from the central naked singularity.
But we are not concerned with this energy as that will be released even if
the collapse leads to the formation of a black hole.
There may be some difference in the two scenarios, but unless the details
are worked out one cannot say anything about it.

Of course, merger of two neutron stars that Singh has suggested will
not be spherically symmetric, and the only reason for considering this
case was that, this is the only case where any significant work has been
done in gravitational collapse leading to naked singularities as well as
the associated quantum processes. The non spherically symmetric situation
has not been studied so far in any detail and we can only discuss
this possibility qualitatively.

It may be argued that if the collapse is not spherically symmetric
the volume of space where the quantum processes operate can increase
as the singularity may be formed along a surface. The same will also be
true for a shell crossing singularity in spherically symmetric
collapse. It is known that the shell crossing singularities in spherically
symmetric collapse are generally weak in some sense (Singh 1996) and hence
one may expect their counterparts in non-spherically symmetric collapse
also to be weak. However, in this discussion we will ignore the strength of
singularity.
If one looks at the arguments presented above for the shell
focusing singularities in spherically
symmetric case it should be clear that the limiting energy does not depend
on the available volume, but only on the admissible distance to which the
quantum effects are expected to dominate. In fact, the expected value is
simply the Planck mass $M_p$
and hence it may not be significantly altered in non-spherically
symmetric situations.
Since there are hardly any calculations of collapse leading to naked
singularities in such situations,
it is difficult to give any firm bound on resulting burst of energy.
In any case, if a substantial fraction of the mass involved
($\approx 10^{33}$ gm) is close to the
singularity surface, the surface must have a linear extent of at least
1 cm, in order to ensure that the singularity will be globally naked.
This arises because if the linear size is much less than the Schwarzschild
radius for the corresponding mass a horizon will almost certainly form.
It may be noted that if the angular momentum exceeds the Kerr limit then
the collapse will not proceed to scales much less than the Schwarzschild
radius.
It will thus require very high level of fine tuning in the initial conditions
to ensure that a sizeable fraction of the available mass is within
$10^{-32}$ cm of the singularity surface which itself is spread over a region
of order of 1 cm. An aspect ratio of $10^{32}$ between two dimensions of
the region
containing substantial mass is impossible to achieve in natural circumstances.
Such a geometry does not appear to be built into the
initial conditions arising from binary neutron stars or a collapse of
single star that Singh has considered. Even a shell crossing singularity
in spherically symmetric collapse will not give rise to such a situation
as in a generic collapse most of the mass will be inside the shell
where density will be generally high.
Thus it appears that the limiting value of energy generated in
non-spherically symmetric case can be obtained by multiplying the spherically
symmetric limit by a `tolerable' aspect ratio. As will be argued in the
following paragraph this aspect ratio is not likely to be very large
and hence the spherically symmetric limit does not need to be increased
significantly when considering non-spherically symmetric case.

The collapse of neutron star binary or collision of two neutron
stars has been studied in great detail
in connection with generation of gravitational waves and some hydrodynamic
calculations of such systems have been performed (e.g., Rasio \& Shapiro 1992;
Centrella \& McMillan 1993, Ruffort et al.~1996; Ruffort \& Janka 1998).
In all cases that
have been studied the end state is found to be nearly spherical, with
aspect ratio much less than 10 between two dimensions.
These models have an extent which is typically few times the Schwarzschild
radius and not much
collapse is required before either a singularity or a horizon is formed.
These calculations
may involve some approximations, but since the discrepancy between the estimated
and observed energy for Gamma-ray bursts is more than 30 orders of magnitude,
more sophisticated calculations may not be required at this stage.
Thus if Singh thinks that such collapse will lead to a density distribution
that is spread along the singularity surface,
then he should repeat these calculations and demonstrate the occurrence of
such density distributions.

There will be additional complication due to presence of angular momentum
in non spherically symmetric collapse. In fact, Singh~(1998) has argued
that angular momentum in certain collapse calculation is too large to form
a Kerr black-hole. However, that does not imply that collapse will lead to
a naked singularity. If the angular momentum is conserved, then
as the collapse proceeds further, rotation will dominate over gravitation
and the star will
become unstable leading to some mass loss along with angular momentum loss.
Alternately, the angular momentum may be lost through gravitational waves.
The collapse cannot continue unless some angular momentum is lost.
If the angular momentum distribution is such that there is not much
angular momentum in the central region, or somehow the angular momentum in
central region is transferred outside, the central region may collapse to
form a singularity, but in that case the situation will not be any different
from situation without angular momentum and the limits obtained earlier will
apply.  If the resulting
system continues to have too much of angular momentum to form a Kerr black-hole
it will continue to loose mass and angular momentum as the collapse
proceeds.
If a black-hole is not formed, then by the time collapse reaches the Planck
length most of the mass will be lost and the naked singularity if at all it
forms will have little matter around it to generate any energy.
If the angular momentum continues to be larger than Kerr limit, the limiting
mass around the singularity will again be of the order of Planck mass.
Thus it appears that the spherically symmetric limit on energy generation
cannot be increased  by invoking angular momentum.
In fact, the presence of angular momentum will work against the formation
of naked singularity, as the collapse to singularity cannot proceed unless
substantial angular momentum is lost from the central region.

We thus conclude that naked singularities coupled with quantum processes
as suggested by Singh~(1998) do not offer a viable explanation
for the Gamma-ray bursts as the likely energy output from such sources
is far below the observed values. In fact the expected energy output
from such a system is probably less than what is emitted in a solar flare
and is unlikely to be of any astrophysical consequence.

\bigskip
\beginsection References

{\parindent=0 pt\everypar{\hangindent=20 pt}

S. Barve, T. P. Singh, C. Vaz, L. Witten, 1998, gr-qc/9805095

P. R. Brady, I. G. Moss, R. C. Myers, 1998, Phys.\ Rev.\ Let.\ 80, 3432.

J. M. Centrella and S. L. W. McMillan, 1993, Astrophys.\ J. 416, 719.

S. K. Chakrabarti and P. S. Joshi, 1994,  Int.\ J.  Mod.\ Phys.\ D3  647.

F. A. Rasio and S. L. Shapiro, 1992, Astrophys.\ J. 401, 226.

M. Ruffort and H.-Th. Janka, 1998, astro-ph/9804132.

M. Ruffort, H.-Th. Janka and G. Schafer, 1996, Astron.\ Astrophys.\ 311, 532

T. P. Singh, 1996, gr-qc/9606016.

T. P. Singh, 1998, gr-qc/9805062.

R. M.  Wald, 1997, gr-qc/9710068.

E. Witten in {\it Quantum gravity and beyond} Eds.\ F.
Mansouri and J. J. Sciano (1992).

}

\end